An Ultrasonic analog for a laser

Richard L Weaver[1], Oleg I Lobkis[1], and Alexey Yamilov[2]

[1]Department of Theoretical and Applied Mechanics, University of Illinois, 104 So Wright Street, Urbana, IL 61801

[2]Department of Physics, University of Missouri-Rolla, 1870 Miner Circle, Rolla, Missouri 65409

Abstract

We report measurements on ultrasonic systems analogous to random lasers. One system entails unstable ultrasonic feedback between distinct transducers, another involves a piezoelectric device that emits spontaneously and by stimulation. Both systems are found to exhibit behaviors similar to those of lasers. Over a wide range of parameters we observe narrow single emission lines, sensitivity to linear cavity properties, complex multi-mode emissions, and line narrowing.

Laser oscillation is not necessarily a quantum phenomenon [1]. Indeed, the observation by Huyghens in 1665 of mode locking between two pendulum clocks may be the first report of locked oscillators. Classical lasers systems [2] have pedagogic value in understanding lasing but have played little role in laser research. Here we report two ultrasonic classical laser systems which we believe will have both pedagogic and research value. Because ultrasonic systems permit probes and controls to a degree not possible in optics, these systems should permit detailed experiments that complement those possible on lasers. One of our uasers [3] is constructed from a reverberant elastic body attached to a piezoelectric sensor which feeds, after suitable amplification, a piezoelectric actuator. (see figure 1a) Another is formed from a single piezoelectric transducer in an operational amplifier auto-oscillator. Each is a random wave system with gain; each exhibits stimulated emission of acoustic waves; each exhibits many of the complex behaviors seen in optical lasers. In both overlapped and non-overlapped modal regimes (i.e. whether the modes of the passive cavity are isolated or not) we observe narrow single frequency emission, thresholds as a function of gain, sensitivity to linear cavity properties, complex multiple frequency emissions, and line narrowing [4]. We observe these behaviors in both ray-chaotic and multiply scattering systems, both of which we term "random."

The systems we present here more closely resemble random lasers than they do conventional highly-directional lasers. Random and ray-chaotic lasers have complex physics, and a potential for practical applications [4-22]. These lasers differ from conventional lasers in that feedback is due to random multiple scattering, or reverberation in a chaotic cavity. Theoretical understanding of lasers is complicated by their essential nonlinearity and by the complex interplay of several times scales; this is especially the case for random lasers. These include Heisenberg times (inverse mean level spacing), Thouless times to diffuse over relevant length scales, absorption times, escape times, Ehrenfest time [at which the correspondence with ray optics breaks down], and rates of gain. Experiments with random lasers are complicated by challenges in precisely controlling properties on the scale of the optical wavelength (e.g. material geometry) or finer, limited information on processes far from the sample boundary, and inability to measure the phase of the electric field [4]. Numerical experiments for realistic 3-d systems are computationally demanding. For these reasons a full understanding of random lasers has been slow to develop, and venues for more precise experiments are to be desired.

Ultrasound has provided useful analog experiments on quantum dots and optical resonators [23,24]. Similarly it may be expected that classical wave experiments such as those presented here could be useful for research in laser dynamics, especially random lasers. These systems have spontaneous and stimulated emission, feedback, gain and saturation. The systems introduced here are not presented as exact analogs to optical lasers, as their saturation dynamics may well differ from that of a laser. We claim that the analogies are nevertheless good, and that the uasers' behavior is similar enough to that of optical lasers, and their design sufficiently simple, that they will complement experiments on optical lasers.

Consider the ultrasonic system pictured in figure 1. A piezoelectric device with useful sensitivity in the range from 100 to 2000 kHz and a diameter of 3 mm is placed at an arbitrary position $\boldsymbol{x}$ on an irregular elastic body. We typically use aluminum (shear and longitudinal wavespeeds 3.1 and 6.15 mm/μsec) of generic ray-chaotic shape and volume between 50 and 5000 $cm^3$. Relevant wavelengths lie in the range between 30 and 1.5 mm. The signal from the sensor at ***x*** is amplified by 40 dB [25], passed (optionally) through a filter to control the frequency range available, sampled by a digitizer, passed through an adjustable voltage divider and amplified by a further 40 dB before being passed to a piezoelectric actuator at ***y*** that provides the ultrasonic source. A third transducer monitors the acoustic state of the elastic block at ***z***. Electronic feedback (familiar from audio acoustics) provides the stimulated emission; the high Q elastic body provides the wave feedback.

With sufficient gain (small enough voltage division) the circuit auto-oscillates with a strength governed by saturation of the amplifiers. Figure 1 shows the behavior in the time domain after the circuit is turned on. Like that of an optical laser, the transition to uasing shows strong threshold-like behavior as a function of gain. For this configuration at gains less than 50 dB there is no auto-oscillation. At 50 dB (upper curve), we observe a slow transition, over about one second, from a wide band low amplitude noisy background to a saturated auto-oscillation at about 700 kHz. At slightly higher gain (lower curve), the transition is quicker and the final state has higher amplitude.

Figure 2a shows the spectrum of the ordinary acoustic response $| H(\omega) |$ of the passive system consisting of transducers and block but no feedback. $H(\omega)$ is the response of the transducer at **y** when an impulse is applied to the transducer at **x**. $H(\omega)$ is the product of the elastodynamic Green's function $G(\omega)$ of the sample[26] and the transducer functions. The broad peaks around 700, 1000 and 1500 kHz are characteristic of these transducers. That the uasing seen above takes place near 700 kHz is thus not surprising; this is in one of the main pass bands of the transducers. Sample steady-state spectra are shown in figures 2b. We observe sharp uaser lines (the frequencies of auto oscillation) within the broad peaks of the ordinary passive spectrum. In optics this would correspond to lasing near the maximum of the gain profile. At modest gains we typically observe a single line, but on occasion, especially at larger gain, the uaser shows many isolated frequencies. The precise position of the line(s) depends on actuator and sensor position, on strength of gain, on temperature, and sometimes hysteretically on the history of these parameters. Line widths are extraordinarily narrow, much narrower than would be expected based on the Q (30,000) of the elastic body at these frequencies. We monitor the frequency of the auto oscillation with a five-digit counter (a 0.1 second integration time at these frequencies). Line positions are observed to be narrow to within our resolution (10 Hz) and stable over periods of minutes, depending only on temperature. This degree of stability is comparable to that obtained using similar circuits and quartz crystal oscillators.

Figure 2c shows a fine scale overlay of the uaser spectrum and the ordinary passive spectrum $|H(\omega)|$. On this scale the ordinary spectrum has features related to the reverberant cavity. At these frequencies the modes of the passive cavity are closely spaced ("overlapped") and cannot be resolved, thus the features in $|H(\omega)|$ are Ericson fluctuations[27]. It may be seen that the uaser oscillation tends to occur at a local maximum of the ordinary spectrum, but not precisely. Nor is the uaser line necessarily near the strongest of the peaks.

Figure 2d shows the spectrum obtained at high gain in a large volume (3000cc) high Q body. The many lines (all narrow to within our resolution of 10 Hz) are tentatively ascribed to nonlinear mixings between at least two primary uaser lines near 750 and 1500 kHz. At these acoustic amplitudes nonlinearity is confined to the amplifiers and their saturation; it is not in the wave propagation.

Theory for the low amplitude linear dynamics of the structure below saturation may be described by

$$\begin{aligned} &S(t) = X(t) \otimes G(x,y,t) \otimes Y(t) \otimes V(t) \\ &H_{xy}(t) = X(t) \otimes G(x,y,t) \otimes Y(t) \\ &V(t) = g(t) \otimes S(t) \end{aligned} \qquad (1)$$

where S(t) is the signal out of the sensor at ***x***, and V(t) is the signal into the actuator at ***y***. They differ by the gain g(t). The symbol ⊗ represents a temporal convolution. X and Y are transducer transfer functions and G is the elastodynamic Greens function [24,26].

Seeking solutions of the form V(t) = exp{iλt}, we find that freely decaying or growing oscillations are the solutions λ to

$$\tilde{H}_{xy}(\lambda) = \int H_{xy}(t)\exp\{-i\lambda t\}dt = \tilde{X}(\lambda)\tilde{G}(\lambda)\tilde{Y}(\lambda) = 1/\tilde{g}(\lambda) \qquad (2)$$

The tilde represents a Fourier transform. At zero gain, eqn(2) implies solutions λ wherever H(ω) is infinite. These λ have positive imaginary part, and correspond to freely decaying natural oscillations. In the uasing regime, i.e, for great enough gain, the fastest growing mode is that solution λ which has the most negative imaginary part.

In figures (3) we plot |H(ω)| and | H(ω) g(ω) –1 | for a gain setting just above the minimum for uasing and for low frequencies as assured by the use of a low-pass filter. Modal density here is 1 mode/kHz at 200 kHz; thus the peaks in | H(ω) | correspond to eigenmodes of the body. When gain is at the threshold for uasing, Im λ ~ 0, and the free

vibration condition (2) reduces to $H(\omega)g(\omega) = 1$. Figure 3a shows $|H|$ itself, and shows the peak centered at 194.106 kHz that appears to be related to the uasing line at 194.095 kHz. (At these low frequencies our frequency resolution is better; it is governed by the one second capture time of our 500000 word digitizer when capturing at 0.5 MSa/sec ). Figure 3b shows $| H g - 1 |$, where g is measured at the frequency of interest. $| H g - 1 |$ has a sharp minimum at the uaser line position. This behavior is also observed at high frequency where modal overlap is large and the fluctuations in the ordinary spectrum are not modes but Ericson fluctuations[27].

Theory (2) predicts $|H g - 1|$ to be zero at the uaser line. We argue that the critical feature in this plot is not only how close the value of $|Hg-1|$ approaches zero, but how quickly it does so as a function of $\omega$. A sharp minimum corresponds to rapid variations of Hg-1 with respect to changes in real $\omega$. But Hg–1 is analytic in $\omega$, so a sharp minimum must correspond to an actual zero at a neighboring complex value of $\omega$. Thus the simple linear theory is supported. A nonlinear model that included saturation would more complex and would require a deeper understanding of the amplifiers; it is not attempted here. A complete exploration of the laser analogy will require such, and comparison with three- or four- level laser equations.

Uasing is found to occur in disordered multiply scattering structures as well as ray-chaotic cavities. We have examined 2-d multiply scattering diffusing and localizing systems [24], and a simple two-room system [24] that localizes at low frequency; behaviors are similar. Some of these structures have higher absorption than do the ballistic bodies and correspondingly require higher gain before they uase.

In a low Q system consisting of a Plexiglas block the lines remain more narrow than we can resolve. At these frequencies the Plexiglas is only moderately reverberant; typical rays reflect only once or twice from the walls before being absorbed. The system therefore better resembles conventional acoustic feedback. Lines in this Plexiglas body are less stable against variations in gain than they are in aluminum.

The above system is recommended in part by its great simplicity and flexibility. That its stimulated emission is non-local and that it exhibits no significant spontaneous emission may, though, preclude some of its usefulness as a laser analog. The standard optical laser consists of local oscillators (atoms) that lock by means of their mutual connection to the radiation field they generate [28]. In the absence of access to a common field, they emit only by spontaneous emission, and do so incoherently. Such a system differs in possibly important ways from the non-local sensor-actuator pair of figure 1. To better model a local stimulated emission of radiation, and better represent spontaneous emission, a different design is called for. Towards that end we built a model of a pumped single atom [29,30] consisting of an Wein bridge[31] operational-amplifier oscillator circuit in which a piezoelectric transducer replaces one of the usual capacitors. See figure (4). In the absence of the cavity, i.e, when the transducer is unconnected to a solid body, the circuit oscillates, but does so in an unsteady fashion, with frequency changing erratically. When contact is made with a high Q elastic body, the oscillation narrows and stabilizes. Uasing is highly sensitive to cavity properties. The presence of uasing, and the frequency of uasing, vary with position on the body, with the choice of body and with changes in the body (e.g. adding absorption.) Thus the oscillation is like that of a single-atom laser, with an oscillator that self-locks by means of its connection to the radiation field it generates.

These points are supported by figure 4. The spontaneous emission spectrum of the isolated Wein bridge including transducer unattached to the elastic body has a finite width of 30 to 1000 Hz. When attached to the elastic body, the line shifts, and narrows to a width finer than our resolution. Equations like (1) and (2) for the linear dynamics of this system, equations that exhibit spontaneous emission, stimulated emission, and mode-locking, are not difficult to derive. Modeling of saturation, and thereby a more complete exploration of the laser analogy, has not been attempted.

It may be noted that the above systems differ from the "sasers" proposed by others [32,33] based on stimulated emission of phonons from defects or of sound from oscillating bubbles.

Optical lasers are useful in practice owing to their propensity for coherent emission, their high intensity, and their potential for rapid switching. These features are of little value in acoustics; coherence is the rule not the exception, intensity is limited by power availability, and maximum speeds are limited by the moderate frequencies. Thus we are reluctant to claim acoustic applications for a uaser. We do express an opinion that such devices could model optical lasers. Because of the ease with which their parameters can be controlled and their state monitored, such devices could be of value in studies of laser dynamics.

Systems such as those of figures 1 and 4 may be modified in various ways chosen to explore issues in optical lasing. An obvious modification would consist of two or more independent sensor-actuator pairs like those of fig(1) applied to the same body.

Another would apply multiple "atoms" like that of figure 4 to a single solid or fluid body of one, two or three dimensions, with any of a variety of boundary conditions. Each transducer would represent one continuously pumped atom. Many questions occur, arising from issues originating in optical systems. To what extent can the analogy be perfected? What behaviors cannot be re-produced? What is the effect of noise on line width? What conditions lead to chaotic uasing? What is the effect of time-varying acoustic parameters? What is the effect of nonlinearity in the acoustics? What is the effect of multiple scattering on the minimum gain needed for uasing? What is the shape of the uasing mode? How does coupling between modes affect the uasing? Is there a mode-locking phase transition [34]? Detailed studies of ultrasonic systems could advance our knowledge of their optical counterparts. We look forward to a period of extensive experimentation.

We thank Andrey Chabanov, Hui Cao, Sri Namachivaya and Lippold Haken for discussions. We particularly thank L Haken for help in the design of the Wein bridge. This work has been supported by the National Science Foundation CMS-0201346.

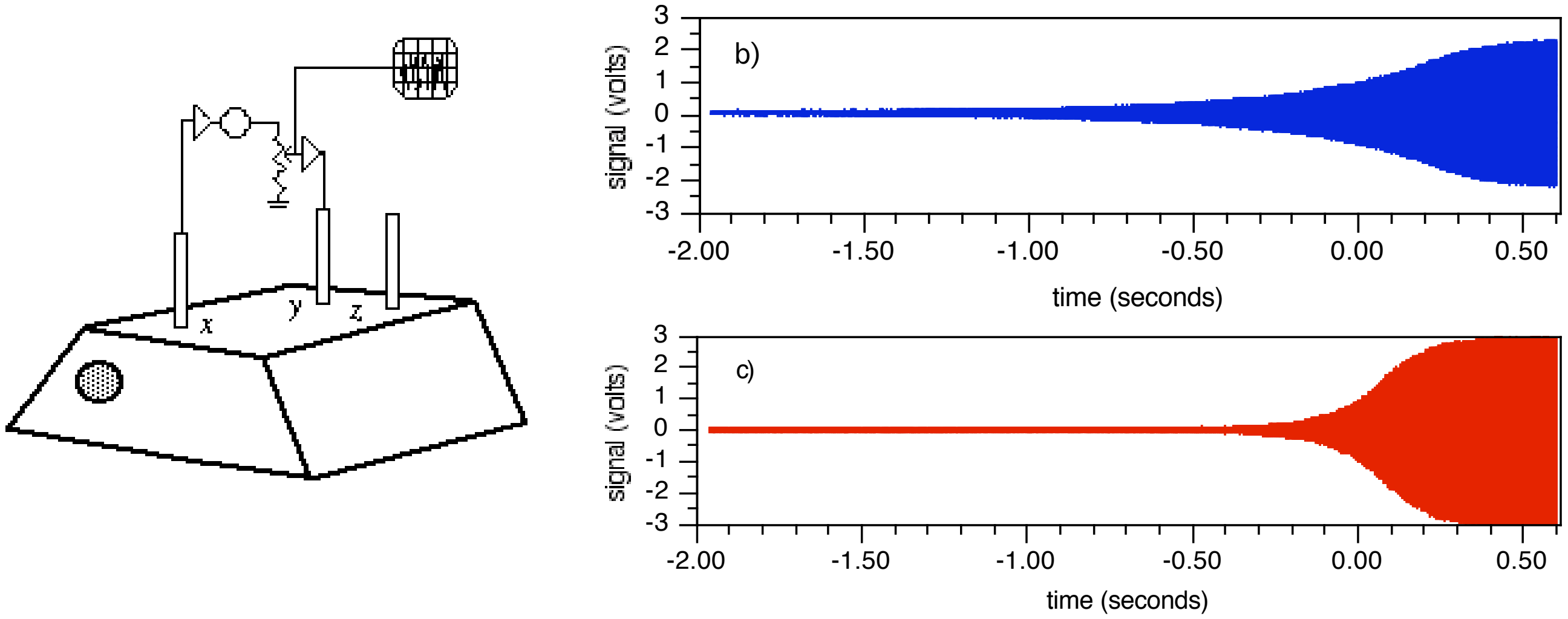

Figure 1] In the simplest uaser, a piezoelectric sensor monitors elastic waves at ***x***; these signals are amplified, optionally filtered, optionally monitored by an oscilloscope, voltage-divided, and amplified again before being fed to a piezoelectric actuator at ***y***. A third transducer monitors system response at *z*. At sufficient gain one observes an unstable growth of a single frequency auto-oscillation. At slightly higher gain (lower curve) the growth is faster.

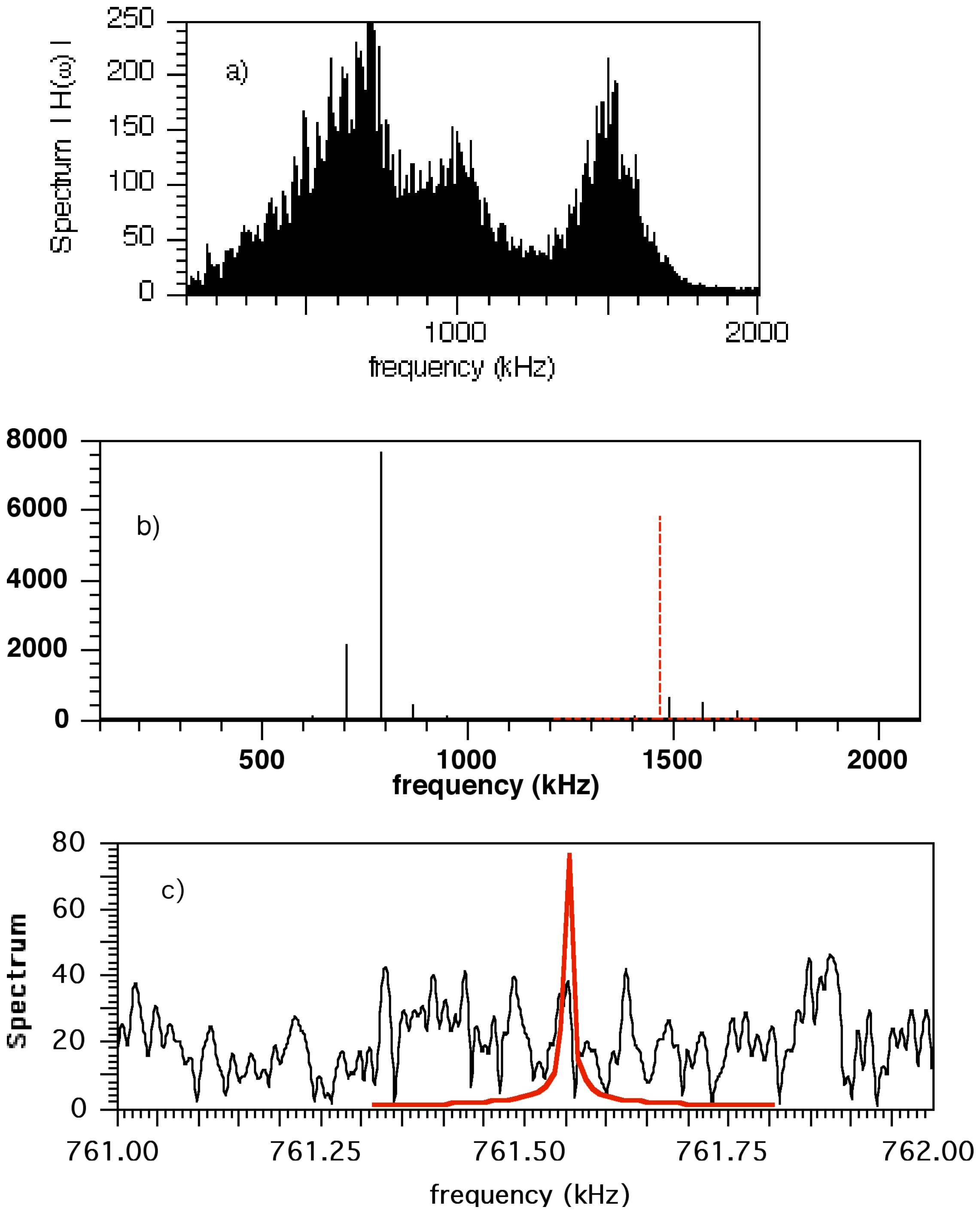

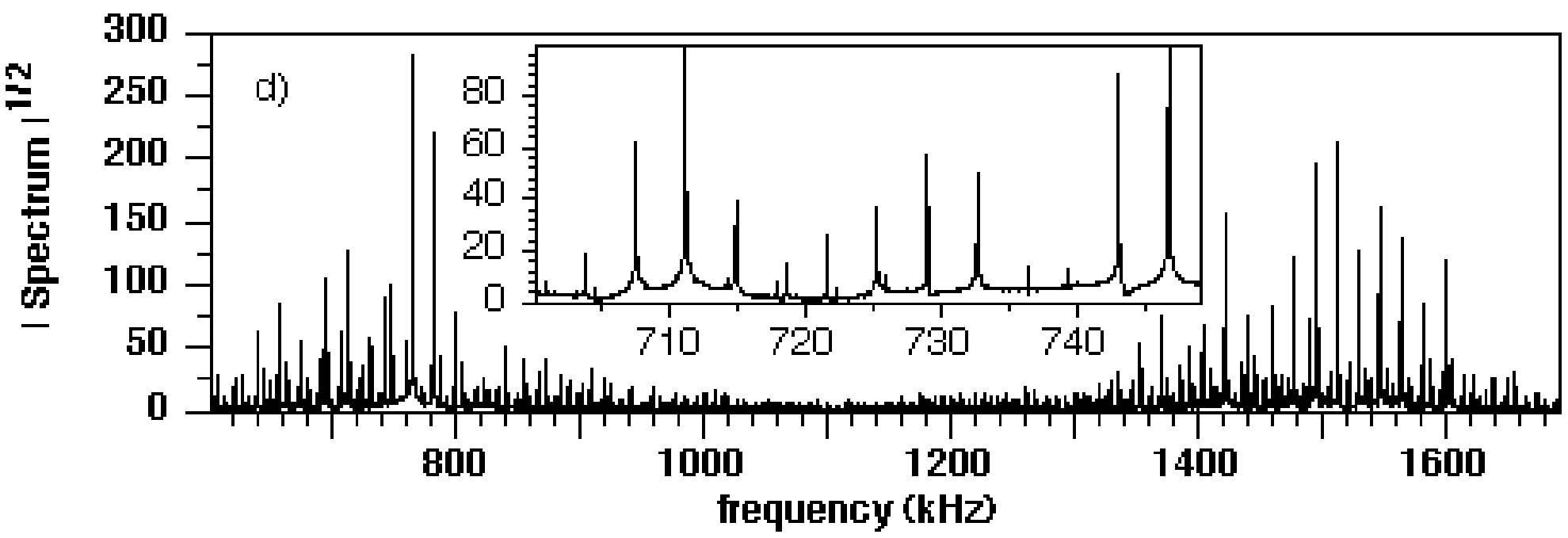

Figure 2] Qualititative behaviors of the uaser. Panel a shows the ordinary passive spectrum of a typical system like that of figure 1, obtained by Fourier analysing the transient response of the transducer at ***y*** to an impulsive forcing at ***x***. The broad peaks at 700, 1000 and 1500 kHz are characteristic of the transducers, as are the finer structures with spacings of the order of 43 kHz (related to a 23 microsecond round trip travel time up and down the transducer barrel.) The finest structures (on a sub kHz scale and not visible here) are due to the elastic block and correspond to a speckle-like diffuse reverberant random response in this ray-chaotic body. Panel b shows two typical uaser spectra consisting of one (dashed) or a few (solid) sharp lines. Panel c shows a fine resolution comparison of the ordinary spectrum and the corresponding uaser spectrum. It can be seen that the uaser line (apparent width 10 Hz is artificially dictated by the amount of time captured) occurs near a local maximum of the ordinary spectrum. Panel d shows the (square root of) the spectrum of a uaser auto oscillation for a case of high gain, a case with a multitude of lines. The inset shows the spectrum at finer resolution. These line spacings are dominated, but not exclusively, by a value of 3.7 kHz, suggesting a near-periodicity of 270 microseconds.

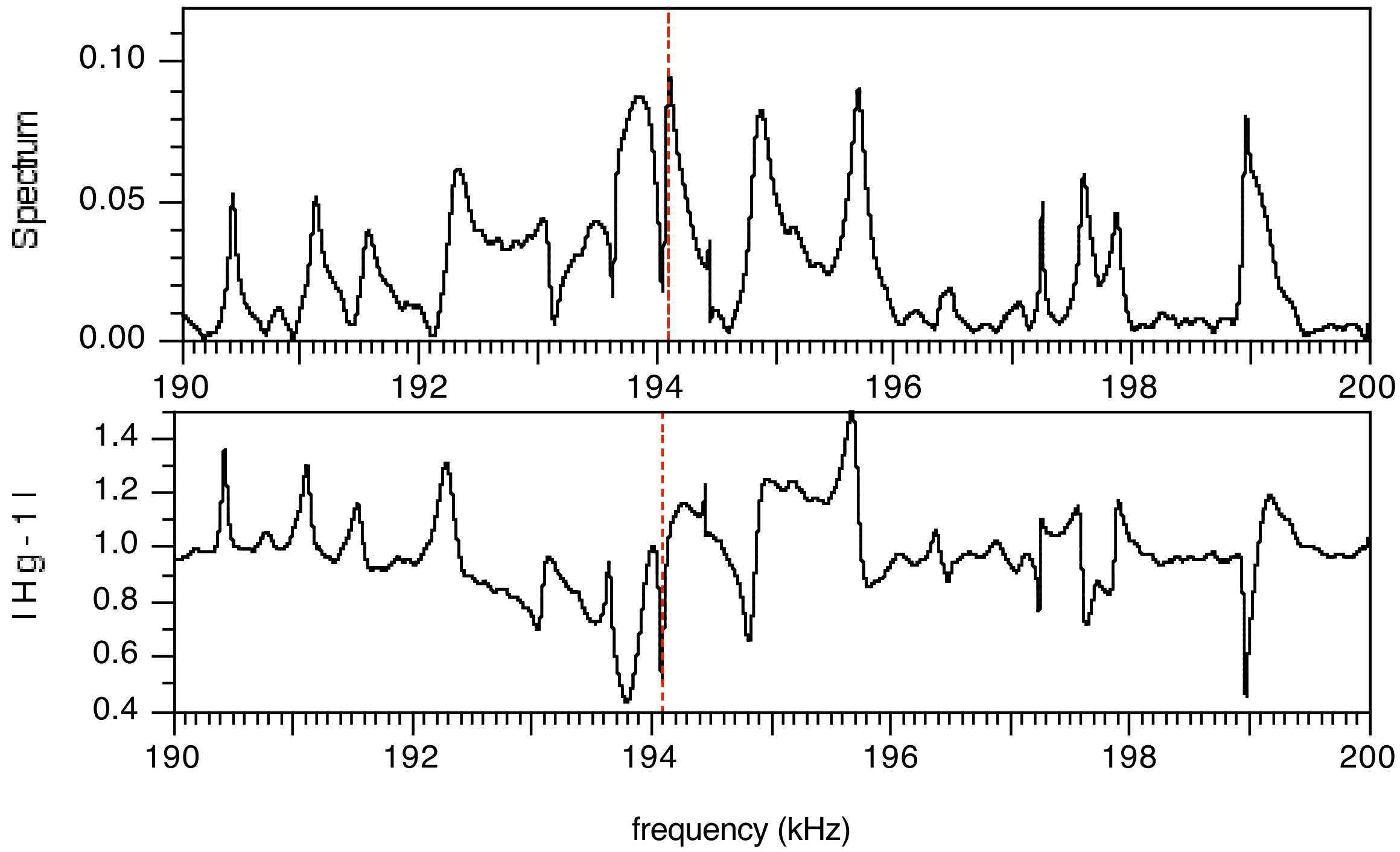

Figure 3] The ordinary spectrum l H l, and uaser line at 194.1 kHz, of a small block at low frequencies (a). The uaser line is close to a maximum of the ordinary spectrum, i.e, close to an eigenmode of the small block. The uaser line position is compared (b) with measured l H g –1 l, showing that the line occurs near a zero of Hg-1.

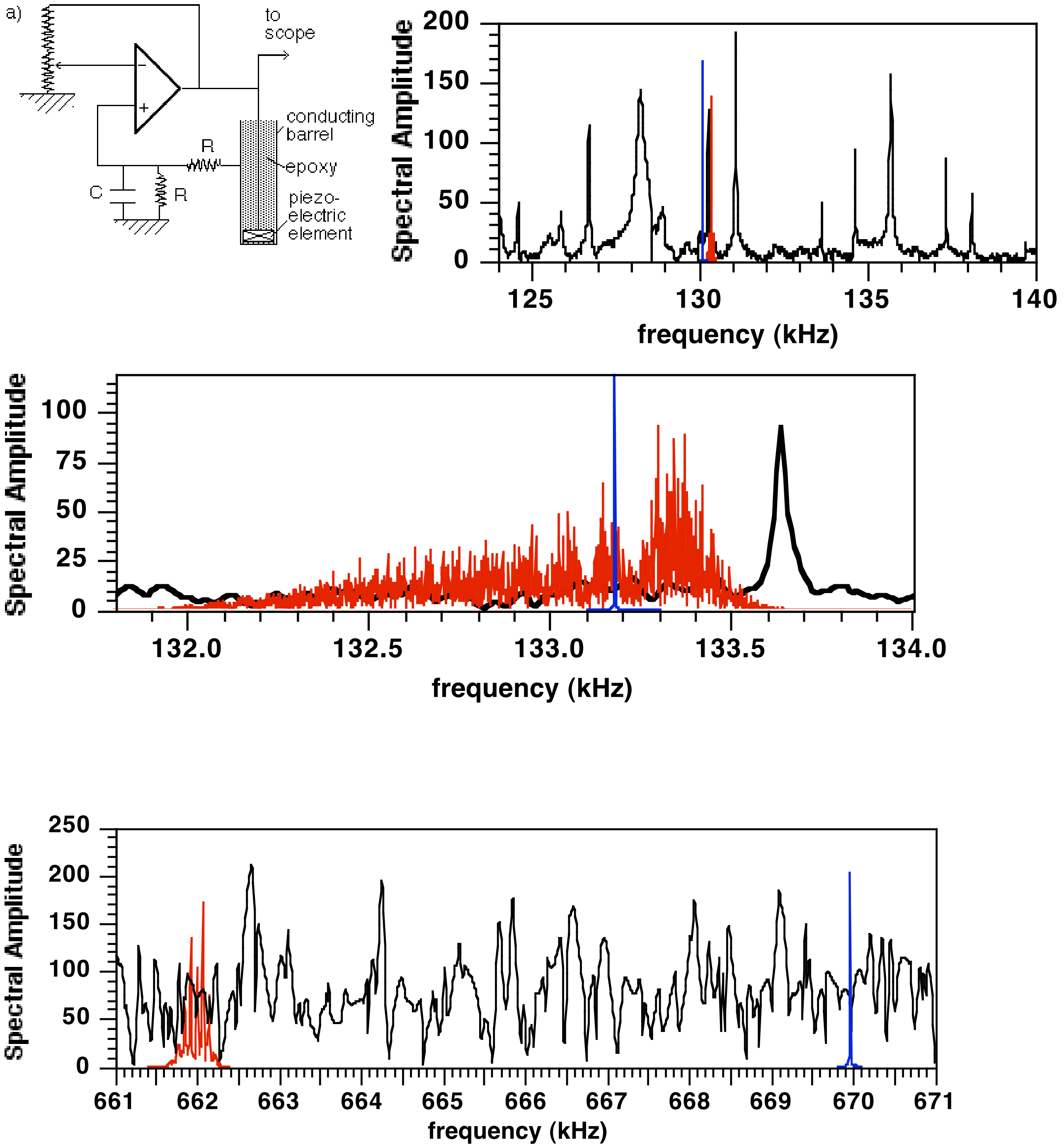

Figure 4] A single atom uaser is formed by inserting a piezoelectric transducer in an operational amplifier circuit, a "Wein bridge" oscillator. The electronic parameters R, C, and the capacitance of the transducer specify the frequency range in which auto-oscillation can occur. In the absence of contact with a reverberant body, the circuit oscillates erratically, spontaneously emitting acoustic radiation. This is illustrated in the plots as a wide noisy grey (red online) curve. When in good contact with a finite reverberant elastic body, the oscillator stabilizes, as indicated in the sharp lines (blue online). The three cases above differ by different position on the block, or different choice of circuit parameter R. Continuous curves indicate the ordinary pulse-echo spectrum of the system when the transducer is disconnected from the Wein bridge and connected to an ultrasonic pulser and receiver.